# Spontaneous RNA condensation into nanoring structures

J. F. Ruiz-Robles, A. M. Longoria-Hernández, N. A. Gerling-Cervantes, L. E. Sanchez-Diaz, E. Reynaga-Hernández, B. I. Ivlev, and J. Ruiz-Garcia

Institute of Physics, Universidad Autónoma de San Luis Potosí, Obregón 64, San Luis Potosí, 78000 México.

The effect of polyvalent cations, such as spermine, on the condensation of DNA into very well defined toroidal shapes have been well studied and understood. However, a great effort has been made trying to obtain similar condensed structures from either ssRNA or dsRNA, which the latter carries similar negative charges as dsDNA, although it adopts a different double helical form. But the analogous condensation of RNA molecules into well-defined structures has so far been elusive. In this work, we show that ssRNA molecules can be easily condensed into well-defined nanorings structures on a mica surface, where each nanoring structure is formed mostly by a single RNA molecule. The condensation occurs in a concentration range of different cations, from monovalent to trivalent. The structures of the RNA nanorings on mica surfaces were observed by atomic force microscopy (AFM). The samples were observed in tapping mode AFM analysis and were prepared by drop evaporation of a solution of RNA in the presence of the different cations. As far as we know, this is the first time that nanorings or any other well-defined condensed RNA structures have been reported in the presence of simple salts. The RNA nanoring formation, can be understood by an energy competition between the hydrogen bonding forming hairpin stems – weakened by the salts– and the hairpin loops. This result may have an important biological relevance, since it has been proposed that RNA is the oldest genome coding molecule and the formation of these structures could have given it stability against degradation in primeval times. Even more, the nanorings structures could have the potential to be used as biosensors and functionalized nanodevices.

Keywords: RNA, nanoring structures, atomic force microscopy, cations.

## INTRODUCTION

Since Watson and Crick discovered the structure of the DNA,[1] its functions inside the cells, viruses or bacteria have been investigated. But in recent decades, scientists' attention has gone to the study of RNA which is the molecule involved in the proteins production,[2] and it has a lot more functions inside the cell than DNA and it is also believed to be the precursor molecule of the origin of life.[3] The condensation of nucleic acids is an issue that has been fascinating scientists of diverse fields. The nucleic acids condensation is defined as the collapse of extended chains of DNA or RNA into highly ordered and compacted particles, which might have only one or a few molecules and that distinguishes it from random aggregation due to the high degree of order that molecules present within the condensed structure. Thus the term condensation can be associated with ordered aggregation of the nucleic acids into a limited size and well defined morphology. Generally, the process of condensation of nucleic acids occurs at the nanometric scale. This phenomenon has been observed in cells,[4] spermatozoids,[5] bacteria[6] and viruses.[7] The nucleic acid condensation is an issue that has direct importance in several research fields, such as cellular biology, virology, therapeutics, nanotechnology and others.

In the last decades, studies have been focusing on trying to obtain RNA condensed structures to compare its behavior with DNA. One of the first observations on DNA condensation is its condensation in the cell´s nucleus, where chromosomes are made up of DNA tightly-wound around histone proteins, which order and package the DNA into structural units known as nucleosomes.[8] DNA condensation into toroidal shapes has been observed inside bacteriophage virus capsids by the interaction with spermine.[9] In addition, it has been discovered that DNA condenses into rod-like and toroidal shapes in the presence of polyvalent cations in solution[10] or even inside of liposomes.[11,12] The experimental evidence shows that DNA condensation occurs when 90% of its charge is neutralized by counterions.[10] There are several forces present in the DNA condensation phenomenon; the loss of entropy promotes the collapse, however, the major contributions are given by the multivalent cations in the solution. The DNA condensates form well-defined structures independently of the DNA size, and these structures are preferably toroids of similar sizes.

Thus, the addition of small amounts of multivalent cations to solutions containing double-stranded DNA leads to inter-DNA attraction and its eventual condensation. Surprisingly, an expected similar condensation is suppressed in double-stranded RNA, which carries a similar negative charges as DNA, but assumes a different double helical form.[13] Thus, highly charged DNA molecules are expected to repel each other, and yet can be condensed by certain multivalent cations into structured aggregates.[10,11,12] The condensation phenomenon is biologically important, since compaction of anionic DNA and RNA molecules by oppositely charged cationic agents enables efficient packaging of genetic material inside living cells and viruses. It has been shown that the effective DNA intramolecular attraction is mainly due to electrostatic contributions in the presence of multivalent cations. However, double-stranded (ds) RNA helices resist condensation under conditions where DNA duplexes condense easily.[14] The condensation of ssRNA by multivalent cations has also been attempted, however, so far there is no evidence of structural condensation of ssRNA. The reason might be that ssRNA in solution presents a secondary structure, that fluctuates. Although one can obtain a minimum energy secondary structure, there are other structures that are very close in energy to the minimum energy secondary structure, see Fig. 1, the ssRNA structure is fluctuating between difference structures. Therefore, to obtain a well defined compacted condensed structure is not expected.

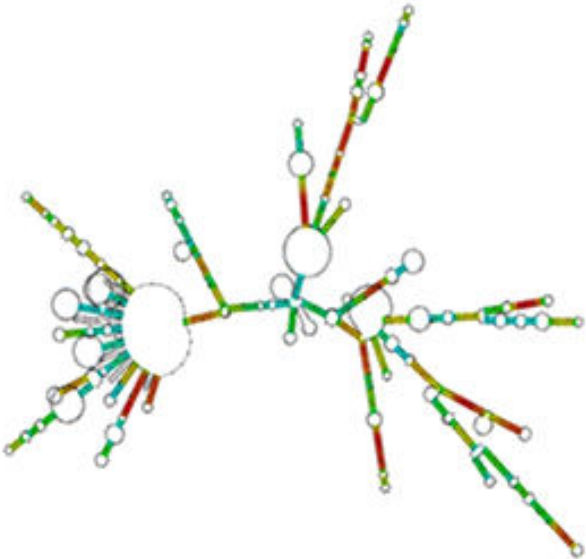

Fig.1. Secondary structure of the RNA that codes for Enhanced Green Fluorescent protein (EGFP) obtained by the Vienna RNA algorithm.

In this work, we used monovalent, divalent and trivalent cations, in the range from 1 to 15 mM. The salts acted as chaotropic agents, helping in the disruption of the ssRNA secondary structure due to the salts weaken the hydrogen bonding. After depositing the RNA solution on the mica surface, the samples were observed by an atomic force microscopy (AFM), which revealed the formation of the RNA nanorings. The nanorings are mostly one molecule thick, but the estimated number of RNA molecules forming the nanorings depends on the valence of the cation. As far as we know, this is the first report on the formation of condensed structures of RNA molecules.

**MATERIALS AND METHODS**

We used five mRNA molecules; four of them were purified from the CCMV virus and the RNA mixture was directly used (RNA1 contain 3,171 nucleotides (nt); RNA2 contain 2,774 nt; RNA3 contain 2,173 nt and RNA4 of 824 nt).[17] The fifth RNA molecule used was a pure mRNA that codes for the EGFP whose sequence is composed of 996 nucleotides and the minimum energy secondary structure is shown in Fig. 1. To prevent RNA degradation, the RNA molecules were kept in a buffer that consists in 0.9 M of NaCl, 0.02 M tris-HCl, 0.01 M EDTA at pH 7.0. In the case of the viral RNA, the purity and concentration were determined by UV-visible measurements (UV-Vis Shimadzu UV-2401). A curve of the sample absorbance was obtained ant it is compared with the characteristic values of the RNA. The purity of the RNA sample is calculated as: P = Abs260/Abs280, where Abs260 is the absorbance at 260 nm, Abs280 is the absorbance at 280 nm and P is the virus purity value.[16] For RNA, the purity is considered good when this ratio is at least 1.5 with a maximum at 260 nm.[19] The virus concentration is obtained by: C = A260 x 40 µg/mL. For the RNA used in this work, the absorbance at 260 nm was 1.75, which means that the concentration of the sample was 70.13 µg/mL and the purity was 2.13 (0.4207/0.1975). After verifying that the RNA purity was very good, the RNA was stored in sterilized vials and was conserved in the freezer at -80 ºC to further avoid RNA degradation.

For the sample preparation, all vials, flasks, pipettes, and even the deionized water were sterilized by autoclaving at 121 ºC for 15 min. Buffer solutions were prepared with three different salts (monovalent, divalent and trivalent) at 1, 5, 10, 12 and 15 mM concentrations for each salt. The salts used were NaCl, $MgCl_2$ and $CeCl_3 \cdot 7H_2O$. Then a 0.07 mg/mL solution of RNA in DNase free water was prepared. 5 µL of RNA solution was aggregated with a sterilized Hamilton syringe at 1 mL of every correspondent salt solution and the vials were agitated softly for RNA dissolution. RNA solutions in the different salts were deposited in freshly cleaved mica and dried at ambient temperature without disturbance.

**AFM CHARACTERIZATION**

Samples were imaged using a Nanoscope III Multimode AFM equipped with a piezoelectric scanner, with a maximum scan range of 10 µm (x and y) and 3.8 µm (z) from VEECO/Digital Instruments. Height, amplitude, and phase images were obtained in Tapping Mode (oscillation frequency 250-300 kHz) in ambient conditions, using cantilevers with a spring constant of 48 N/m (MikroMash, Watsonville, CA, USA) with nominal radius of curvature <10 nm and the scan rate was 1-3 Hz. The samples were usually imaged after drying the sample, but some samples were imaged within a few days after preparation. Samples were stored at 4 °C and some samples were imaged a month after preparation to check for any changes in morphology.

## RESULTS AND DISCUSSION

The RNA samples were deposited on freshly cleaved mica, which is a silicate that is widely used as substrate for the deposition of biomolecules to be observed by AFM, due to it is molecularly flat and smooth surface. The typical roughness observed on the mica surface is about 0.09 nm.

However, when the surface of the freshly cleaved mica comes into contact with a water solution, the hydrated potassium ions desorb, causing the surface of the mica to acquire a negative charge.

Since RNA in water also acquires a high negative charge, it is important to have positive ions in the RNA buffer solution. Thus the positive ions in the solution have two important effects in our experiments; the first one is to neutralize or even reverse the mica surface charge to a positive surface that facilitate the absorption of the RNA molecules. The second effect is that they act as chaotropic agents, as discussed below.

Nanoring structures were observed after a few µL of the RNA in solution with cationic ions is deposited on a recently cleaved mica surface, as is shown in Figs. 2 and 3. The mica surface in contact with water develops negative charges on its surface. In addition, the RNA molecules are also highly negative molecules. Thus, the cations from the salt not only neutralize the mica surface charge but in the case of multivalent cations can reverse the mica surface charge and increase the adhesion of the nucleic acid. In addition, we use a low concentration of salts to prevent the formation of salt crystals on the mica surface.[20] The nanoring structures are very well defined, and it call the attention that most nanoring structures are made mostly by one molecule. This can be explained by a wetting effect that can break-down due to a capillary instabilities, where the adsorbed film decomposes into a collection of segregated, non-overlapping molecules.[21] It is important to have a pure RNA without salts in the precipitation-purification process.

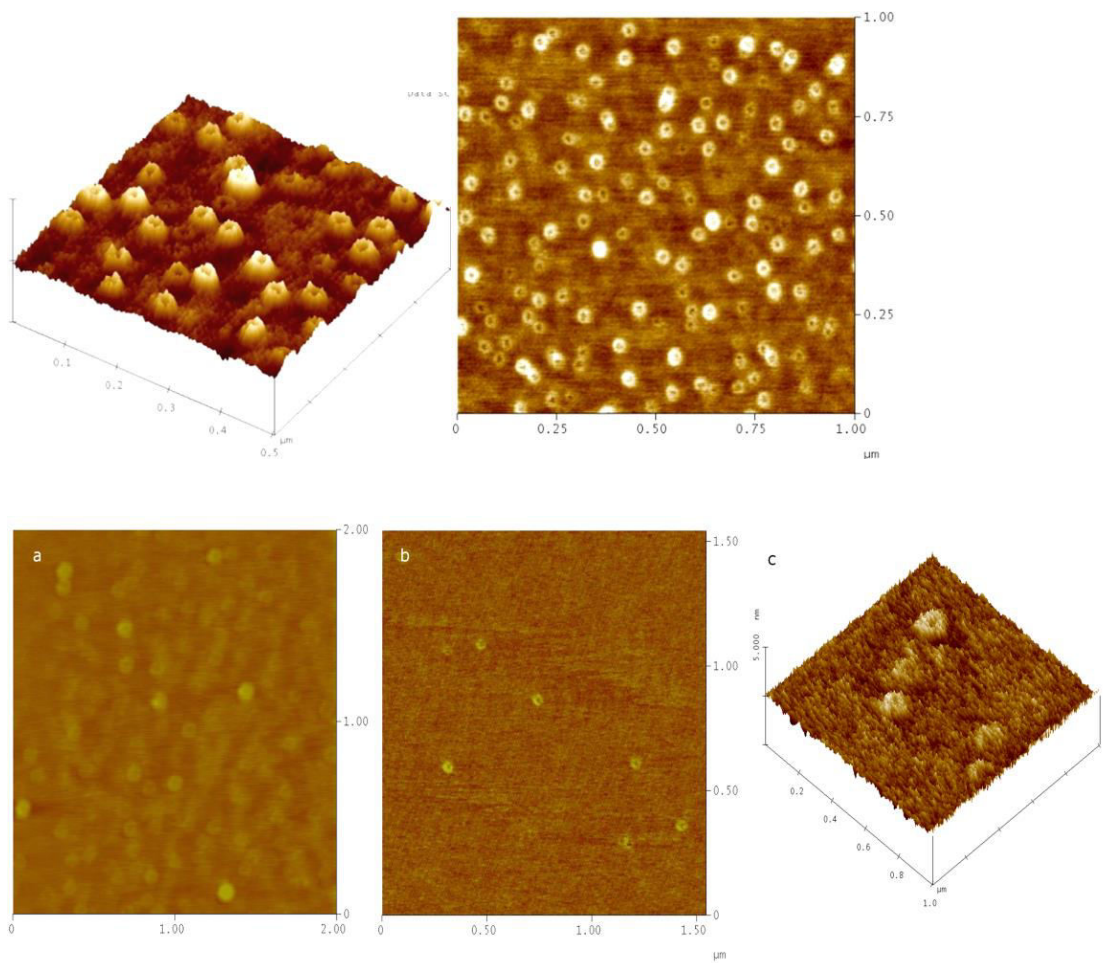

Fig. 2. AFM images of CCMV RNA mixture forming nanorings observed in the presence of different cations with an external diameter of: (a). 12 mM of $Na^+$, (b) 12 mM of $Mg^{2+}$, (c) 5 mM $Ce^{3+}$. The external diameter of the nanorings. The scan size of the images is a) 2x2 µm²; b) 1.5x1.5 µm² and c). 1.0x1.0 µm².

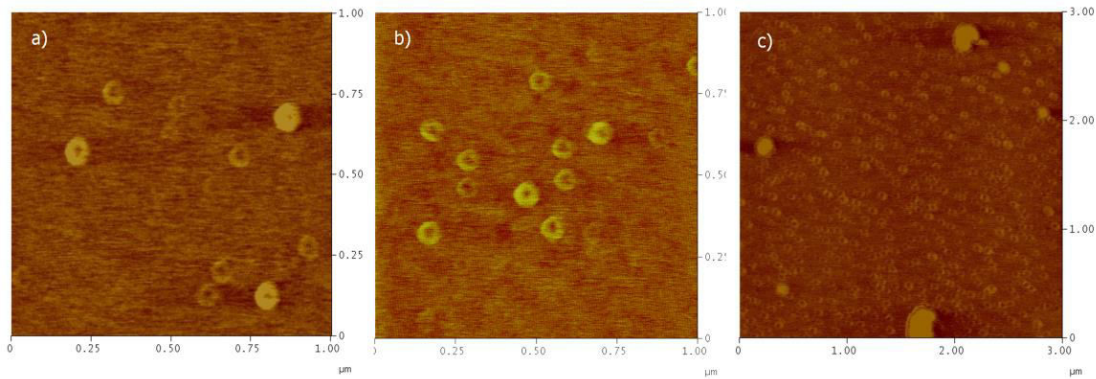

Fig. 3. AFM images of nanorings obtained using the EGFP RNA observed in the present of different cations or concentrations. (a) 10 mM of $Mg^{2+}$, (b) 15 mM of $Mg^{2+}$, (c) 5 mM of $Na^{+5}$.

In Table 1, we show the sizes of the RNA nanorings under the different conditions where the experiments were performed. Note that the external diameter of the nanorings ($D_{ext}$) for the monovalent cationic salt is very similar at all concentrations used, which is also of similar size for the divalent cationic salt at the two lower concentrations. Using the thickness of the nanoring and the external and internal diameter, we estimated that these nanorings are approximately made of one mRNA molecule. However, when the concentration of the divalent cationic salt increases, the size of the nanorings also increases. While the size of the nanorings are larger with the trivalent salt even at low concentrations. Also the thickness of the nanorings, for the monovalent and divalent (at low concentration) cationic salts corresponds to thickness of one mRNA molecule (~ 1 nm). However, the thickness increases as the concentration of increases, specially with the divalent and trivalent cationic salts. Even more, at higher concentrations of salts, the nanorings turn into a globular or spherical shape, but yet well separated one from the other, as shown in Fig. 4.

| Ion Salt | + Na | | | | | 2+ Mg | | | | | 3+ Cs | |
|---|---|---|---|---|---|---|---|---|---|---|---|---|
| Concentration, mM | 1 | 5 | 10 | 12 | 15 | 1 | 5 | 10 | 12 | 15 | 1 | 5 |
| Dext, nm | 48 | 56 | 47 | 43 | 58 | 51 | 48 | 82 | 74 | 72 | 63 | 77 |
| Molecules/Nanoring | 1.2 | 1.6 | 1.2 | 0.9 | 1.7 | 1.4 | 1.2 | 3.4 | 2.8 | 2.7 | 2.1 | 3.0 |
| Nanoring Height | 1.1 | 2.6 | 1.5 | 1.1 | 1.2 | 0.9 | 1.2 | 2.33 | 1.5 | 2.2 | 4.3 | 4.4 |

Table 1.

It is worth to remark that DNA condensation is only observed in the presence of multivalent ions in solution, while here we report the RNA condensation in the presence of monovalent to trivalent cations. This condensation can be understood

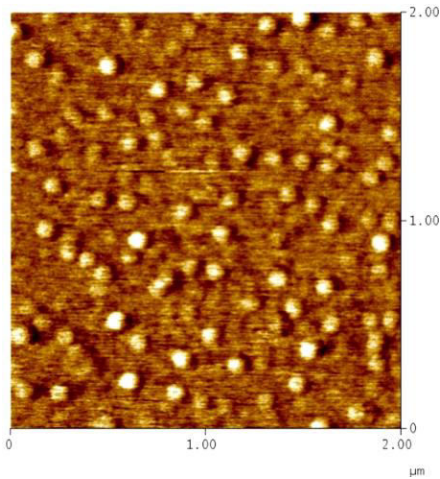

Fig. 4. Globules or spherical shape RNA structures observed at 20 mM of $Mg^{2+}$ cationic salt. The diameter of these structures is 102.66 ± 5.5 nm an their average height is 1.4 ± 0.15 nm

as a competition of the energy related with the two types of hairpins that form the RNA secondary structure, as they are shown schematically in Fig. 5.

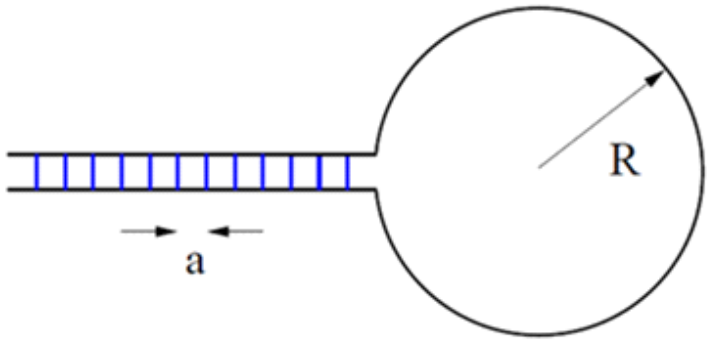

Fig. 5. A schematic short piece of the ssRNA secondary structure showing the formation of a hairpin stem and a hairpin loop. The hairpin stem on the left part contains hydrogen bonds separated by the distance a ~1 nm, due to base pairing when read in opposite directions. Whereas R is the radius of the hairpin loop.

The left part in Fig. 5 is a usual one where two strands are bound by hydrogen bonds, separated by the distance $a$ and it is commonly called a hairpin stem. The energy related to the hairpin stem can be described by the following expression:

$$E_{hyd} = -V_{hyd} \frac{L - 2\pi nR}{2a} \quad (1)$$

where L is the total length of RNA and is the energy gain due to the formation of individual hydrogen bonds, and is the average number of hairpin loops formed in the RNA secondary structure.

In the hairpin loop part, the hydrogen bonds are not involved. Instead there is an elastic energy pay due to bending of the RNA chain. This elastic energy can be evaluated from general arguments. Let us blowup the part of the bent chain as shown in Fig. 6.

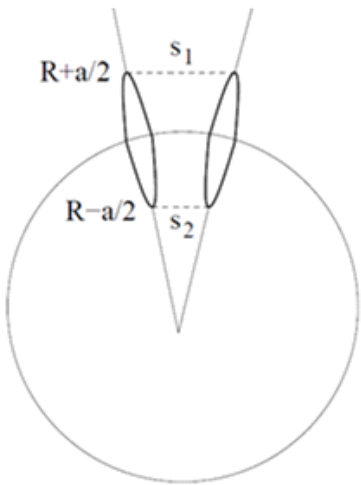

Fig. 6: Schematic figure of the part of the bent chain of nucleotides. The distance between two nucleotides along the hairpin loop is $a$.

The center to center average distance between nucleotides can be estimated as:

$$s_{1,2} = a \pm \frac{a^2}{2R} \tag{2}$$

When $R$ is infinity (no bending), the position is equilibrium related to a minimal covalent energy. The deviation from that equilibrium value (elastic energy $E_{el}$) is quadratic with respect to the displacement $a^2/2R$. It can be written in the form

$$E_{el} = \alpha \frac{V_{cov}}{a^2} \left(\frac{a^2}{2R}\right)^2 \frac{2\pi R}{a} \tag{3}$$

Where $\alpha$ is a numerical parameter and the last factor accounts for the total number of sites along the circle. $V_{cov}$ is the individual covalent energy of one site. As a result, the elastic energy of the circle can be written as:

$$E_{el} = \frac{\pi \alpha}{2} V_{cov} \frac{a}{R} \tag{4}$$

The total energy, $E_{hyd} + E_{el}$, can minimized with respect to , in order to obtain the equilibrium circle radius; the result is,

$$R = a \sqrt{\frac{\alpha V_{cov}}{2 V_{hyd}}} \tag{5}$$

Since the energy of the hydrogen bond is much lower than the energy of the covalent bond and, in addition, gets even weaker in the presence of the chaotropic salts, then the radius of the RNA circle (Nanoring) is larger than the distance a = 1 nm between the subsequent elements of RNA. Since the energy of the C-C bond is of the order of C–C 347 KJ/mol and the hydrogen bonds are of the order of 10-40 KJ/mol, then in this case we can take the lower value due to the present of the chaotropic salts, in order to estimate the value of the constant.

Taking an average diameter value of the nanorings of ~ 60 nm, the value for the constant is of the order of 207.5.

**CONCLUSIONS**

In this work, we show for the first time the formation of condensed structures in ssRNA in the form of nanorings. The formation of these RNA nanostructures can be understood by the energy minimization of the RNA molecules; this is, the competition between the elastic energy due to bending of the RNA chain in the hairpin loop and the hydrogen bonding in the hairpin stem, where the hydrogen bond energy is weakened by the interaction with the chaotropic salts. Also, the concentration of salts is important, since without or at relatively high concentrations of salts, the formation of RNA nanorings is not observed. On the other hand, if the salt concentration is high, the formation of RNA globules is observed instead of nanorings. It has also been recognized that DNA internal structure plays an important role in its toroidal condensation. In the other hand, RNA has many secondary structures, because these structures only differ very little with the secondary structure of the minimum global energy. Therefore, the RNA structure is fluctuating among different secondary structures. Attempts have been made to obtain well-defined condensed structures of both types of molecules, ssRNA and dsRNA, without success. Even though dsRNA has a similar number of charges to dsDNA, no well-defined structures has been observed. As far as we now, our work is the first one to present well defined condensed structures of ssRNA. The observation of the RNA nanoring formation, could lead to a better understanding on the stability of RNA molecules, especially in primeval times, since it is believe that RNA was the first genetic molecule that gave origin to live.

**ACKNOWLEDGEMENTS**

We acknowledge support from the Consejo Nacional de Ciencia y Tecnología (CONACYT-Mexico) through Grants FC-341, CB-254981 and CB-237439. J. R.-G., also acknowledge partial support from the UASLP through the Fondos Concurrentes program.